\documentclass[11pt]{article}
\usepackage{amsfonts}
\usepackage{amssymb}
\newsymbol\contra 1379
\def\a{\alpha}
\def\b{\beta}
\def\g{\gamma}
\def\d{\delta}

\def\o{\omega}
\def\p{\partial}

\def\vn{\varnothing}
\def\vp{\varphi}

\def\th{\theta}
\def\bl{\backslash}
\def\hra{\hookrightarrow}
\def\lra{\longrightarrow}
\newcommand{\ca}{{\mathcal A}}

\newcommand{\cf}{{\mathcal F}}
\newcommand{\cg}{{\mathcal G}}
\newcommand{\ch}{{\mathcal H}}

\newcommand{\cm}{{\mathcal M}}
\newcommand{\cn}{{\mathcal N}}

\newcommand{\ct}{{\mathcal T}}

\newcommand{\U}{{\mathcal U}}
\newcommand{\C}{\mathbb C}

\newcommand{\Ad}{{\mathop{\mbox{Ad}}\nolimits}}
\newcommand{\ad}{{\mathop{\mbox{ad}}\nolimits}}

\newcommand{\id}{{\mathop{\mbox{id}}\nolimits\,}}
\newcommand{\im}{{\mathop{\mbox{Im}}\nolimits\,}}
\newcommand{\Ker}{{\mathop{\mbox{Ker}}\nolimits\,}}
\newcommand{\Tr}{{\mathop{\mbox{Tr}}\nolimits\,}}

\newcommand{\fm}{{\mathfrak M}}
\newcommand{\fc}{{\mathfrak C}}

\newcommand{\fh}{{\mathfrak H}}

\newcommand{\fg}{{\mathfrak g}}

\newcommand{\frs}{{\mathfrak s}}
\newcommand{\fs}{{\mathfrak S}}
\newcommand{\fu}{{\mathfrak U}}
\newcommand{\fx}{{\mathfrak X}}

\begin{document}

\begin{flushright}
 hep-th/0002120
\end{flushright}

\vskip 3cm

\begin{center}
{\LARGE \bf Dressing Symmetries of Holomorphic BF Theories$^*$}
\vskip 1.5cm
{\Large T.A.Ivanova$^1$
and  A.D.Popov$^2$}

\vskip 1cm

{\em Bogoliubov Laboratory of Theoretical Physics\\
JINR, 141980 Dubna, Moscow Region, Russia}\\

\vskip 1.1cm
\end{center}

\begin{center}
{\bf Abstract}
\end{center}

\begin{quote}

We consider holomorphic BF theories, their solutions and symmetries.
The equivalence of \v{C}ech and Dolbeault descriptions of holomorphic
bundles is used to develop a method for calculating hidden (nonlocal)
symmetries of holomorphic BF theories. A special cohomological
symmetry group and its action on the solution space are described.

\vspace{0.5cm}

\end{quote}

\vfill
\hrule width 5.cm

\vskip 3mm

{\small \noindent ${}^{*\ {}}$
supported in part by the grant RFBR-99-01-01076

\vskip 1mm

\noindent ${}^{1\ {}}$
E-mail: ita@thsun1.jinr.ru

\vskip 1mm

\noindent ${}^{2\ {}}$
E-mail: popov@thsun1.jinr.ru}

\newpage

\section{Introduction}
Let $Z$ be a complex $n$-dimensional manifold, $G$ a (complex)
semisimple matrix Lie group, $\fg$ its Lie algebra, $P$ a principal
$G$-bundle over $Z$, $A$ a connection 1-form on $P$ and $F_A=
dA+A\wedge A$ its curvature. Consider the following action:
$$
S_{\mathrm{hBF}}=\int_Z \Tr(B\wedge  F^{0,2}_A),
\eqno(1.1)
$$
where $B$ is an $\ad P$-valued $(n,n-2)$-form on $Z$, $\ad P:=
P\times_G\fg$,  and  $F^{0,2}_A$ is the (0,2)-component of the
curvature tensor $F_A$. The field equations for the action (1.1)
are
$$
\bar\p A^{0,1}+ A^{0,1}\wedge A^{0,1}=0,
\eqno(1.2a)
$$
$$
\bar\p B+ A^{0,1}\wedge B- B\wedge A^{0,1}=0,
\eqno(1.2b)
$$
where $\bar\p$ is the (0,1)-part of the exterior derivative $d=\p +\bar\p$
and $A^{0,1}$ is the (0,1)-component of a connection 1-form $A=A^{1,0}+
A^{0,1}$ on $P$.

\smallskip

Notice that Eqs.(1.2a) coincide with the compatibility conditions
$F_A^{0,2}=\bar\p^2_A=0$ of Eqs.(1.2b), $\bar\p_A=\bar\p + A^{0,1}$.
It follows from Eqs.(1.2) that models (1.1) describe holomorphic
structures $\bar\p_A$ on bundles over complex $n$-manifolds $Z$
and $\bar\p_A$-closed $\ad P$-valued $(n,n-2)$-forms $B$ on $Z$.
So, theories with the action (1.1) generalize topological BF
theories~\cite{Ho,BT} which give a field-theoretic description of
flat connections on bundles over real $n$-manifolds.
Models (1.1) can also be considered as a generalization of
holomorphic Chern-Simons-Witten theories~\cite{Wi,BCOV} defined in
three complex dimensions. Theories with the action (1.1) have been
introduced in~\cite{Po1} and called holomorphic BF theories. They
can be useful in describing invariants of complex manifolds (Ray-Singer
holomorphic torsion and the others) and compactified configurations
in superstring theory~\cite{Po1}. We believe  holomorphic BF theories
(hBF) deserve further developing.

\smallskip

The purpose of the present paper is to describe a procedure of
constructing solutions to Eqs.(1.2) and mappings of solutions into one
another (dressing transformations). We describe a parametrization of
solutions to Eqs.(1.2) by transition functions of topologically
trivial holomorphic bundles and elements of Dolbeault cohomology
groups. We show that all (dressing) symmetries of Eqs.(1.2) can be
calculated with the help of homological algebra methods.

\section{Field equations of hBF theories and their solutions}
\subsection{Flat (0,1)-connections and functional matrix equations}

We consider a complex $n$-manifold $Z$, a principal $G$-bundle $P$ over
$Z$ and a connection
1-form $A$ on $P$. The  curvature $F_A=dA+A\wedge A$ of a connection
$A$ splits into components,
$$ F_A=F^{2,0}_A+ F^{1,1}_A + F^{0,2}_A, $$
and the $(0,2)$-component of the curvature tensor is
$$F^{0,2}_A=\bar\p^2_A=\bar\partial A^{0,1}+A^{0,1}\wedge A^{0,1}.$$
The (0,1)-component $A^{0,1}$ of a connection 1-form $A=A^{1,0}+A^{0,1}$
will be called the (0,1)-connection.

\smallskip

{}For simplicity we shall consider a trivial $G$-bundle $P_0\simeq Z
\times G$. Then for the adjoint bundle $\ad P_0=P_0\times_G\fg$ we have
$\ad P_0\simeq Z\times \fg$, where $\fg$ is the Lie algebra of a group $G$.
Generalization to the case of nontrivial bundles is straightforward and
not difficult. We denote by
$$ \Omega^{p,q}(Z,\fg)$$
the space of $\fg$-valued smooth $(p,q)$-forms on $Z$. Taking a form
$B\in \Omega^{n,n-2}(Z,\fg)$, we introduce the action (1.1) of holomorphic
BF theory and consider field equations (1.2).

\smallskip

Solutions of Eqs.(1.2a) are {\it flat} $(0,1)$-connections $A^{0,1}$.
They can be described in different ways. To show this, let us fix a
covering $\fu =\{U_\a\}$ of a complex manifold $Z$, $\a\in I$. Then consider
a manifold
$$
\U^{(0)}\equiv Z=\mathop{\bigcup}_{\a\in I}U_\a ,
\eqno(2.1a)
$$
and the subsets
$$\U^{(1)}= \mathop{\bigcup}_{\a ,\b\in I}U_\a\cap U_\b,
\eqno(2.1b)
$$
$$
\U^{(2)}= \mathop{\bigcup}_{\a ,\b ,\g\in I}U_\a\cap U_\b\cap U_\g
\eqno(2.1c)
$$
of a manifold $Z$. The summation in (2.1b) and (2.1c)
is carried out over $\a ,\b ,...$ for which $U_\a\cap U_\b\ne\vn$  and
$U_\a\cap U_\b\cap U_\g\ne\vn$.

Let us consider a collection $\psi =\{\psi_\a\}$ of smooth $G$-valued
functions $\psi_\a$ defined on $U_\a$, a collection $f=\{f_{\a\b}\}$
of holomorphic matrices on nonempty intersections $U_\a\cap U_\b$
and suppose that $\psi_\a$'s satisfy the differential equations
$$ (\bar\partial \psi_\a )\psi^{-1}_\a =
(\bar\partial \psi_\b )\psi^{-1}_\b
\eqno(2.2)
$$
defined on $\U^{(1)}$, and $f_{\a\b}$'s satisfy the functional equations
$$
f_{\a\b|\g}f_{\b\g|\a}f_{\g\a|\b}=1
\eqno(2.3a)
$$
on $\U^{(2)}$. Here $f_{\a\b|\g}$ means the restriction of $f_{\a\b}$
to an open set $U_\a\cap U_\b\cap U_\g$. Now, let us define a map of solutions
$\{\psi_\a\}$ of Eqs.(2.2) into solutions $\{f_{\a\b}\}$ of Eqs.(2.3a) by
the formula $\eta : \{\psi_\a\}\mapsto \{\psi_\a^{-1}\psi_\b\}$, and denote
by $\cf$ the subset of those solutions $\{f_{\a\b}\}$ of Eqs.(2.3a) for
which there exists a collection $\psi =\{\psi_\a\}$ of smooth $G$-valued
functions $\psi_\a$ on $U_\a$ such that
$$
f_{\a\b}=\psi^{-1}_\a\psi_\b .
\eqno(2.3b)
$$
Using equations $\bar\p f_{\a\b}=0$, one can easily show that these
$\{\psi_\a\}$ will satisfy Eqs.(2.2) and therefore $\cf =\im\eta$.

\smallskip

Denote by $\fx$ the space of solutions to Eqs.(2.2) and consider a map
$$
\eta :\ \fx\to\cf
\eqno(2.4a)
$$
given for any $\psi =\{\psi_\a\}\in\fx$ by the formula
$$
\eta (\psi )=\eta (\{\psi_\a\})=\{\psi_\a^{-1}\psi_\b\}=f.
\eqno(2.4b)
$$
It is easy to see that if $g=\{g_\a\}$ is an element of the gauge group
$\cg$, then $\psi =\{\psi_\a\}\in\fx$ and
$g^{-1}\psi =\{g^{-1}_\a\psi_\a\}\in\fx$ are projected by the map (2.4)
into the same solution $f=\eta (\psi )$ of Eqs.(2.3). Therefore the
space $\cf$  of solutions to functional equations (2.3) can be identified
with the space of orbits of the group $\cg$ in the set $\fx$,
$$
\cf\simeq \fx /\cg .
\eqno(2.5)
$$

Now consider a $(0,1)$-connection $A^{0,1}$ on $P_0$, restrict $A^{0,1}$ to
$U_\a$'s and consider a collection $A^{0,1}=\{A^{(\a )}\}$ of
(0,1)-connections $A^{(\a )}:=A^{0,1}|_{U_\a}$. In terms
of $A^{(\a )}$'s, Eqs.(1.2a) have the form
$$ \bar\partial A^{(\a )}+ A^{(\a )}\wedge A^{(\a )}=0
\eqno(2.6a)
$$
$$
A^{(\a )}=A^{(\b )} \ \mbox{on}\ U_\a \cap U_\b .
\eqno(2.6b)
$$
Denote by $\cn$ the space of solutions to Eqs.(2.6) and define a map
$$\bar\d^0:\ \fx\to\cn \eqno(2.7a) $$
given by the formula
$$\bar\d^0(\psi )=      \psi\bar\p \psi^{-1}=\{ \psi_\a \bar\p
\psi^{-1}_\a\}=\{A^{(\a )}\}=A^{0,1},
\eqno(2.7b) $$
where $\psi =\{\psi_\a\}\in\fx$.
It is clear that $\{A^{(\a )}\}=\{\psi_\a \bar\p\psi^{-1}_\a\}$
satisfy Eqs.(2.6) if $\{\psi_\a\}$ satisfy Eqs.(2.2).

\smallskip

It is not difficult to see that for any $\{\psi_\a\}\in\fx$ we have
$\{\psi_\a h^{-1}_\a\}\in\fx$ if $h_\a$'s are holomorphic $G$-valued
functions on $U_\a$'s,
$$\bar\p h_\a =0.  \eqno(2.8)$$
Moreover, such $\{\psi_\a\}$, $\{\psi_\a h^{-1}_\a\}\in\fx$  are mapped by
$\bar\d^0$ into the same flat (0,1)-connection $A^{0,1}=\{A^{(\a
)}\}= \{\psi_\a\bar\p\psi^{-1}_\a\}$. We denote by $\fh$ the set of
all collections $h=\{h_\a\}$ of $G$-valued locally defined
holomorphic functions $h_\a : U_\a\to G$. The set $\fh$ is a group
under the pointwise multiplication: $h\chi =\{h_\a\chi_\a\}$ for
$h=\{h_\a\}, \chi=\{\chi_\a\}\in\fh$.  So, it follows from (2.7) that the
space $\cn$ of flat $(0,1)$-connections can be identified with the space of
orbits of the group $\fh$ in the space $\fx$ of solutions to
Eqs.(2.2),
$$ \cn\simeq\fh\bl\fx .  \eqno(2.9) $$

We see that if we find $\psi =\{\psi_\a\}$ from Eqs.(2.2) or Eqs.(2.3),
then we can obtain a flat (0,1)-connection $A^{0,1}=\psi\bar\p\psi^{-1}$
with the help of the map (2.7). Thus, flat (0,1)-connections can be found
not only by solving Eqs.(1.2a) on $\U^{(0)}$ but also by solving Eqs.(2.2)
on $\U^{(1)}$ or Eqs.(2.3a) on $\U^{(2)}$.

\subsection{Moduli space of holomorphic structures}

Consider a (trivial) $G$-bundle $P_0$ over a complex $n$-manifold $Z$
and a (0,1)-connection $A^{0,1}$ on $P_0$. The gauge group $\cg$ acts on
$A^{0,1}=\{A^{(\a )}\}$  by the formula
$$
A^{0,1}\mapsto \Ad_{g^{-1}}A^{0,1}= g^{-1}A^{0,1}g +g^{-1}\bar\p g
=\{g^{-1}_\a A^{(\a )}g_\a +g^{-1}_\a\bar\p g_\a \},
\eqno(2.10)
$$
where $g=\{g_\a\}\in\cg$. Equations (1.2a) and (2.6) are invariant under
the transformations (2.10).

\smallskip

We denote by $\cm$ the set of orbits of the gauge group $\cg$ in the set
$\cn$ of solutions to Eqs.(2.6),
$$\cm =\cn /\cg .
\eqno(2.11)
$$
By definition, $\cm$ is the moduli space of flat $(0,1)$-connections
$A^{0,1}$ parametrizing holomorphic structures $\bar\p_A$ on the bundle
$P_0$. By introducing a projection
$$\pi :\ \cn\to\cm ,
\eqno(2.12)
$$
we obtain a composite map
$$\pi\circ\bar\delta^0 : \
\fx\stackrel{ \bar\d^0}{\lra}\cn\stackrel{\pi}{\lra}\cm \eqno(2.13)
$$ of $\fx$ onto $\cm$.

\smallskip

Recall that on the space $\fx$ we have an action not only of the
group $\cg$ but also of the group $\fh$,
$$
\fh\ni h=\{h_\a\}:\quad\psi\mapsto\psi h^{-1}=\{\psi_\a
h^{-1}_\a\}, \eqno(2.14)
$$
where $\psi =\{\psi_\a\}\in\fx $. This action induces the following
action of $\fh$ on matrices $f_{\a\b}=\psi^{-1}_\a\psi_\b$:
$$
\fh\ni h=\{h_\a\}:\ f_{\a\b}\mapsto\tilde
f_{\a\b}=h_\a f_{\a\b}h_{\b}^{-1}.
\eqno(2.15) $$
Therefore one can introduce the space $\fh\bl\cf$ of orbits of the
group $\fh$ in the space $\cf$ of solutions to Eqs.(2.3). Then,
using the bijection (2.5), we obtain
$$ \fh\bl\cf\simeq \fh\bl\fx /\cg .
\eqno(2.16) $$
By definition, $\fh\bl\cf$ is the moduli space of solutions to functional
equations (2.3). Comparing (2.9), (2.11) and (2.16), we obtain bijections
$$
\cm\simeq \cn /\cg  \simeq \fh\bl\fx /\cg \simeq \fh\bl\cf ,
\eqno(2.17)
$$
i.e. there is a one-to-one correspondence between the moduli spaces of
solutions to Eqs.(2.3) and Eqs.(2.6). We identify these moduli spaces
with the moduli space $\cm$ of holomorphic structures on the bundle
$P_0\to Z$.

\smallskip

Let us denote by $p$ a projection
$$p:\ \cf\to\cm .
\eqno(2.18)
$$
Combining (2.4) and (2.18), we obtain a composite map
$$
p\circ\eta :\ \fx\stackrel{\eta}{\lra}\cf\stackrel{p}{\lra}\cm
\eqno(2.19)
$$
of $\fx$ onto $\cm$ (cf.(2.13)).

\smallskip

Comparing all the maps described above, we obtain the following
commutative diagram:
\newpage
$$\fx$$
$$\bar\d^0\swarrow\quad\searrow\eta$$
$$\cn\qquad\qquad\cf\eqno(2.20)$$
$$\pi\searrow \quad\swarrow p$$
$$\cm$$
Here $\fx$ is the solution space of differential equations (2.2),
$\cn$ is the space of flat $(0,1)$-connections on $P_0$, $\cf$ is the
solution space of functional equations (2.3), and $\cm$ is the moduli space
of holomorphic structures on the bundle $P_0\to Z$.

\subsection{Non-Abelian cohomology and holomorphic bundles}

Results of Sect.2.2 can be reformulated in terms of homological algebra
using sheaves of non-Abelian groups. Namely, let $\fs$ be the sheaf
of germs of smooth $G$-valued functions on $Z$, $\ch$ its subsheaf of
holomorphic $G$-valued functions and $\ca^{0,1}$ the sheaf of flat
(0,1)-connections on $P_0$ (germs of solutions to Eqs.(1.2a)).
We fix a covering $\fu =\{U_\a\}$ of a manifold $Z$ and introduce
the following sets: the set $C^0(\fu ,\fs)$ of 0-cochains of the
covering $\fu$ with values in $\fs$, the set $Z^0(\fu ,\fs )$ of
0-cocycles with values in $\fs$, the set $C^1(\fu ,\fs )$ of 1-cochains
with values in $\fs$, the set $Z^1(\fu ,\fs )$ of 1-cocycles
of the covering $\fu$ with values in the sheaf $\fs$
and the 1-cohomology set $H^1(\fu ,\fs )$.
These sets contain the subsets  $C^0(\fu ,\ch)$, $Z^0(\fu ,\ch )$,
$C^1(\fu ,\ch )$, $Z^1(\fu ,\ch )$ and $H^1(\fu ,\ch )$, respectively.
All the definitions can be found e.g. in~\cite{F, D, O, Hir}.

\smallskip

Recall that by definition $H^0(Z ,\fs )=\Gamma (Z,\fs )=$ $Z^0(\fu ,\fs ),
H^0(Z ,\ca^{0,1})=$ $\Gamma (Z ,\ca^{0,1})=$ $Z^0(\fu ,\ca^{0,1})$.
Moreover, one can always choose a covering $\fu =\{U_\a\}$ such that
it will be $H^1(\fu ,\fs )=H^1(Z ,\fs )$, $H^1(\fu ,\ch )=H^1(Z ,\ch )$.
This is realized, for instance, when $U_\a$'s are Stein manifolds
and we suppose that the chosen covering satisfies the above conditions.
In cohomological terms some of spaces and groups introduced
earlier are defined as follows:
$$
\fh =C^0(\fu ,\ch),
\eqno(2.21a)$$
$$
\cn =H^0(Z ,\ca^{0,1}),
\eqno(2.21b)$$
$$
\cg = H^0(Z ,\fs ),
\eqno(2.21c)$$
$$
\cm\simeq\cn /\cg = H^0(Z ,\ca^{0,1})/H^0(Z ,\fs ).
\eqno(2.21d)$$
Notice also that the space $\fx$ is a subset of the set $C^0(\fu ,\fs )$,
and the space $\cf$ is a subset of the set $Z^1(\fu ,\ch )$. Namely,
$\cf$ is the set of those 1-cocycles $f\in Z^1(\fu ,\ch )$ that are
smoothly equivalent to the cocycle $f_0=\{\mbox{id}_{U_\a\cap U_\b}\}$.

\smallskip

We denote by $i:\ \ch\to\fs$ an embedding of $\ch$ into $\fs$ and define
a map $\bar\delta^0:\ \fs\to\ca^{0,1}$ given for any open set $U$ of the
space $Z$ by the formula
$$
\bar\d^0(\psi_U):= \psi_U\bar\p\psi^{-1}_U,
\eqno(2.22)
$$
where $\psi_U\in \Gamma (U,\fs )$ is a smooth $G$-valued function on $U$.
Let us also introduce an operator $\bar\delta^1$ acting on (0,1)-connections
$A^{0,1}$ by the formula
$$
\bar\delta^1(A^{0,1}):=\bar\p A^{0,1}+A^{0,1}\wedge A^{0,1}.
\eqno(2.23a)
$$
By definition, $\bar\delta^1$ maps any flat (0,1)-connection into zero
and therefore
$$
\bar\delta^1(\ca^{0,1})=0\ \Leftrightarrow\ \ca^{0,1}=\Ker\bar\delta^1.
\eqno(2.23b)
$$

\smallskip

Remember that locally Eqs.(1.2a) are solved trivially, and on any
sufficiently small open set $U\subset Z$ we have $A_U^{0,1}=\psi_U\bar\p
\psi_U^{-1}$, where $A_U^{0,1}\in\Gamma (U,\ca^{0,1})$ and $\psi_U\in \Gamma
(U,\fs )$ is a smooth $G$-valued function on $U$ (cf.(2.22)).
It is easy to see that
$$
A^{0,1}_U=\psi_U\bar\p \psi_U^{-1}=  (\psi_Uh_U^{-1})\bar\p
(\psi_Uh_U^{-1})^{-1},
\eqno(2.24)$$
where $h_U\in \Gamma (U, \ch )$ is an arbitrary holomorphic $G$-valued
function on $U$ (a section of the sheaf $\ch$ over $U$). Therefore,
the sheaf of germs of solutions to Eqs.(1.2a) is isomorphic to the
quotient sheaf $\ch\bl\fs$. Notice that the (left) action of the sheaf $\ch$
on $\fs$ is described for any open set $U$ by the formula $\psi_U\mapsto
\psi_U h^{-1}_U$, where $\psi_U\in \Gamma (U, \fs )$,
$h_U\in \Gamma (U, \ch )$. Thus, we have the exact sequence of sheaves
$$
{e}\lra\ch\stackrel{i}\lra \fs\stackrel{\bar\d^0}\lra
\ca^{0,1} \stackrel{\bar\d^1}\lra  e,
\eqno(2.25)
$$
where $e$ is a marked element of the considered sets (the identity in
the sheaf
$\ch\subset\fs$ and zero in the sheaf $\ca^{0,1}$). From (2.25) we obtain
the exact sequence of cohomology sets~\cite{F,D,O},
$$
e\lra H^0(Z,\ch)\stackrel{i_*}\lra H^0(Z,\fs)\stackrel{\bar\d^0_*}\lra
H^0(Z,\ca^{0,1}) \stackrel{\bar\d^1_*}\lra H^1(Z,\ch)
\stackrel{\rho}\lra H^1(Z,\fs),
\eqno(2.26)
$$
where the map $\rho$ coincides with the canonical embedding induced by
the embedding of sheaves $i: \ch\to \fs$.

\smallskip

By definition the 1-cohomology sets $H^1(Z,\ch)$ and $H^1(Z,\fs)$
parametrize the sets of equivalence classes of holomorphic and
smooth $G$-bundles over $Z$, respectively.
The kernel $\Ker\rho =\rho^{-1}(e)$ of the map $\rho$ coincides
with a subset of equivalence classes of topologically trivial
holomorphic bundles $P$. Therefore we have
$$
\fh\bl\cf =\Ker\rho ,
\eqno(2.27)
$$
where the space $\fh\bl\cf $ is the moduli space of solutions to Eqs.(2.3).
By virtue of the exactness of the sequence (2.26), the space $\Ker\rho =
\fh\bl\cf $ is bijective to the quotient space (2.21d).
So, the bijections (2.17) follow from the exact sequence (2.26) and
we have
$$
\cm\simeq H^0(Z, \ca^{0,1})/H^0(Z,\fs )\simeq\Ker\rho  .
\eqno(2.28)
$$
Recall that $\cm$ is the moduli space of holomorphic structures on the
bundle $P_0$.

\subsection{Algebra-valued forms and Dolbeault cohomology}

Now let us consider Eqs.(1.2b) on globally defined $\fg$-valued
$(n,n-2)$-forms $B$ on $Z$. These equations can be rewritten in the form
$$\bar\p_AB=0,\eqno(2.29)$$ and Eqs.(1.2a) coincide with the compatibility
conditions $F^{0,2}_A=\bar\p^2_A=0$ of Eqs.(2.29).

The gauge group $\cg$ acts on a field $B\in\Omega^{n,n-2}(Z,\fg)$
by the formula
$$ B\mapsto \Ad_{g^{-1}}B=g^{-1}Bg,
\eqno(2.30)
$$ where $g\in\cg$. Notice that the action (1.1) and Eqs.(1.2)
are invariant under the gauge transformations (2.10), (2.30) and
under the following ``cohomological" symmetry transformations:
$$ B\mapsto B+\bar\partial_A\Phi , \eqno(2.31) $$ where
$\Phi\in\Omega^{n,n-3}(Z,\fg)$. By virtue of this invariance,
solutions $B$ and $B+\bar\p_A\Phi$ of Eqs.(1.2b) are considered
as equivalent.

\smallskip

Equations (1.2b) are linear in $B$. For any fixed flat (0,1)-connection
$A^{0,1}$ the space of nontrivial solutions to Eqs.(1.2b) is the $(n,n-2)$th
Dolbeault cohomology group
$$
H^{n,n-2}_{\bar\partial_A;P_0}(Z):=\frac{\{B\in\Omega^{n,n-2}(Z,\fg ):
\bar\partial_A B=0\}}{\{B=\bar\partial_A\Phi , \Phi\in \Omega^{n,n-3}
(Z,\fg )\}}.
\eqno(2.32)
$$
So, the space of nontrivial solutions to Eqs.(1.2b) forms the vector
space $H^{n,n-2}_{\bar\partial_A;P_0}(Z)$ depending on a solution
$A^{0,1}$ of Eqs.(1.2a).

\smallskip

{}For a fixed flat (0,1)-connection $A^{0,1}=\psi\bar\p\psi^{-1}=
\{\psi_\a\bar\p\psi^{-1}_\a\}$ any solution of Eqs.(1.2b) has the form
$$ B=\psi B_0\psi^{-1}=\{ \psi_\a B_0^{(\a )}\psi^{-1}_\a \}=\{B^{(\a )}\},
\eqno(2.33) $$
where $B_0=\{B_0^{(\a )}\}$ is an arbitrary solution of the equations
$$\bar\p B_0=0.\eqno(2.34a)$$
Here $B^{(\a )}:=B|_{U_\a}$ and $B_0^{(\a )}:=B_0|_{U_\a}$ are restrictions
of $B$ and $B_0$ to an open set $U_\a$ from the covering $\fu =\{U_\a\}$,
$\a\in I$. On nonempty intersections $U_\a\cap U_\b$ we have $B^{(\a )}=
B^{(\b)}$. These compatibility conditions for $B=\{B^{(\a )}\}$ lead to
the following compatibility conditions for $B_0=\{B_0^{(\a )}\}$:
$$ B_0^{(\a )}=f_{\a\b}B^{(\b )}_0 f_{\a\b}^{-1}, \eqno(2.34b)$$ where
$f_{\a\b}:=\psi^{-1}_\a\psi_\b$, and $\{\psi_\a\}$ satisfy
Eqs.(2.2). It follows from Eqs.(2.2) that $f_{\a\b}$'s
are holomorphic matrices on $U_\a\cap U_\b$. Therefore, $\{f_{\a\b}\}$
can be chosen as transition functions in a holomorphic bundle $P$.

\smallskip

Notice that the space of nontrivial solutions to Eqs.(2.34) is the
standard Dolbeault cohomology group $$ H^{n, n-2}_{\bar\p ; P}(Z)=
\frac{\{\bar\p\mbox{-closed}\ \ad P\mbox{-valued}\
(n,n-2)\mbox{-forms\ on}\ Z\}}{\{\bar\p\mbox{-exact}\ \ad
P\mbox{-valued}\ (n,n-2)\mbox{-forms\ on}\ Z\}}, \eqno(2.35)$$ where
$P$ and $\ad P=P\times_G\fg$ are holomorphic bundles defined by
transition functions $\{f_{\a\b}\}=\{\psi_\a^{-1}\psi_\b\}$.  Formula
(2.33) defines an isomorphism of the vector spaces (2.32) and (2.35).

\smallskip

To sum up, one can easily construct solutions of Eqs.(1.2b) if
one knows solutions of Eqs.(1.2a) and Eqs.(2.34). Moreover, the
space of solutions to Eqs.(1.2) forms a vector bundle $\ct\to\cn$,
the base space of which is the space $\cn$ of solutions to Eqs.(1.2a),
and fibres of the bundle $\ct$ at points $A^{0,1}\in\cn$ are the
vector spaces $H^{n,n-2}_{\bar\p_A;P_0}(Z)$ of nontrivial solutions to
Eqs.(1.2b). Recall that the gauge group $\cg$ acts on solutions
$(A^{0,1},B)$ of Eqs.(1.2) by formulae (2.10), (2.30). Therefore,
identifying points $(A^{0,1},B)\in \ct$ and $(g^{-1}A^{0,1}g +
g^{-1}\bar\p g, g^{-1}Bg)\in\ct$ for any $g\in\cg$, we obtain
the moduli space
$$ \fm=\ct /\cg \eqno(2.36) $$ of solutions to Eqs.(1.2). The space $\fm$
is a vector bundle over the moduli space $\cm$ of flat
$(0,1)$-connections with fibres at points $[A^{0,1}]\in\cm$
isomorphic to the Dolbeault cohomology groups (2.32).

\section{Dressing transformations in hBF theories}
\subsection{Cohomological symmetry groups}

In Sect.2 we have discussed the correspondence between flat
(0,1)-connections $A^{0,1}=\{A^{(\a )}\}$ on a $G$-bundle $P_0\to Z$
and 1-cocycles $f=\{f_{\a\b}\}$ defining topologically trivial
holomorphic bundles $P$ over $Z$. It follows from this correspondence
that if we define an action of some group on the space $\cf$ of
transition functions $f$ of topologically trivial holomorphic
bundles, then using the correspondence $A^{0,1}\leftrightarrow f$
(see the diagram (2.20)), we obtain an induced action of this group on
the space $\cn$ of flat (0,1)-connections on the bundle $P_0$. In
this section we introduce a {\it special cohomological group} and describe
its action on the space $\cf$.

\smallskip

Consider a collection $h=\{h_{\a\b}\}\in C^1(\fu,\ch )$ of holomorphic
matrices such that
$$h_{\a\b |\g}=h_{\a\g |\b},\eqno(3.1)$$
where $h_{\a\b |\g}$ means the restriction of $h_{\a\b}$ to an open
set $U_\a\cap U_\b\cap U_\g$. The constraints (3.1) are not severe.
They simply mean that sections $h_{\a\b}\in\Gamma (U_\a\cap U_\b ,\ch )$
of the sheaf $\ch$ over $U_\a\cap U_\b$ can be extended to sections
of the sheaf $\ch$ over the open set
$$
\U^{(1)}= \mathop{\bigcup}_{\a ,\b\in I}U_\a\cap U_\b ,
$$
where the summation is carried out in all $\a ,\b\in I$ for which
$U_\a\cap U_\b\ne\vn$. In other words, it follows from (3.1) that there
exists a holomorphic map
$$
h_{\U^{(1)}}:\quad  \U^{(1)}\to G
\eqno(3.2a)$$
such that
$$h_{\a\b}=h_{\U^{(1)}|U_\a\cap U_\b}. \eqno(3.2b)$$
One can identify $h=\{h_{\a\b}\}=\{h_{\U^{(1)}|U_\a\cap U_\b}\}$ and
$h_{\U^{(1)}}$. Such $h\in\Gamma (\U^{(1)},\ch )$ form a subgroup
$$
\bar C^1(\fu ,\ch):=\{h\in C^1(\fu ,\ch): h_{\a\b |\g}=h_{\a\g|\b}\
\mbox{on} \ U_\a\cap U_\b\cap U_\g\ne\vn\}
\eqno(3.3)
$$
of the group $C^1(\fu,\ch)$.

\smallskip

We consider  $\bar C^1(\fu ,\ch)$ as a {\it local } group,
i.e. we choose a neighbourhood $\fc$ of the identity $e$ in
$\bar C^1(\fu ,\ch)$ and take elements $h$ only from $\fc\subset
\bar C^1(\fu ,\ch)$. The local group $\fc$ is a representative
of the germ of the group $\bar C^1(\fu ,\ch)$ at the point $e\in
\bar C^1(\fu ,\ch)$. We define the following action of the group
$\fc$ on the space $\cf$:
$$
T(h,\ .\ ):\  f_{\a\b}\mapsto f^h_{\a\b}= T(h, f)_{\a\b}=
h_{\a\b} f_{\a\b}h^{-1}_{\b\a},
\eqno(3.4a)
$$
where $h\in \fc$, and a 1-cocycle $f=\{f_{\a\b}\}=
\{\psi^{-1}_\a\psi_\b\}\in\cf$ defines a topologically trivial
holomorphic bundle $P$. It is easy to see that $\{f^h_{\a\b}\}$
satisfy Eqs.(2.3a) by virtue of the definition (3.3)
of the group $\bar C^1(\fu ,\ch)$ and therefore $\{f^h_{\a\b}\}$
is a 1-cocycle. Moreover, for $h\in\fc\subset\bar C^1(\fu ,\ch)$
there exists a 0-cochain $\psi^h=\{\psi^h_\a\}\in\fx\subset C^0(\fu ,\fs )$
such that
$$
f^h_{\a\b}=  (\psi_\a^h)^{-1}\psi_\b^h,
\eqno(3.4b)$$
since small enough deformations do not change the topological
trivializability of holomorphic bundles. This well-known statement
follows from the equality $H^1(Z,\frs_P)=0$~\cite{GrH}, where $\frs_P$
is the sheaf of smooth sections of the bundle $\ad P$.

\smallskip

So, we have a map
$$ T:\ \fc\times\cf\to\cf ,
\eqno(3.5)
$$
and to each $h\in\fc$ there corresponds a bijective transformation
$$T_h:\ f\mapsto T(h,f)\eqno(3.6)$$
of the set $\cf$. The map $t:\ h\mapsto T_h$ is a homomorphism of the
group  $\fc$ into the group $Bij(\cf )$ of all bijective
transformations of the set $\cf$. Notice that maps (3.6) are
connected with maps between bundles $(P, f)$ and $(P^h, f^h)$, where
a bundle $P^h$ is defined by  transition functions
$\{f^h_{\a\b}\}$. These bundles are diffeomorphic but not
biholomorphic. A diffeomorphism of $P$ onto $P^h$ is defined by a
0-cochain $\psi^{-1}\psi^h=\{\psi^{-1}_\a\psi^h_\a\}\in C^0(\fu,\fs
)$.  Moreover, the bundles $P$ and $P^h$ become biholomorphic after
the restriction to $\U^{(1)}:\ P_{|\U^{(1)}}\simeq P^h_{|\U^{(1)}}$.
In other words, a map $T_h:\ f\mapsto f^h,\ h\in\fc$, defines a {\it
local} biholomorphism $P_{|\U^{(1)}}\to P^h_{|\U^{(1)}}$ which does not
extend up to the biholomorphism of $P$ and $P^h$ as holomorphic
bundles over $Z$.

\smallskip

More general transformations of the space $\cf$ of topologically
trivial holomorphic bundles  can be found by discarding the
conditions (3.1) on matrices $\{h_{\a\b}\}$. Namely, let us consider
the transformations (3.4a) with an arbitrary element $h=\{h_{\a\b}\}$ of
the group $C^1(\fu ,\ch )$. Then consider the equations
$$
h_{\a\b } f_{\a\b} h_{\b\a }^{-1}\ h_{\b\g} f_{\b\g} h_{\g\b }^{-1}
h_{\g\a} f_{\g\a}h_{\a\g}^{-1}=1
\eqno(3.7)
$$
on $\{h_{\a\b}\}$.
These equations mean that $f^h=T(h,f)=\{h_{\a\b}f_{\a\b}h^{-1}_{\b\a}\}$
is a 1-cocycle. For each solution $h=\{h_{\a\b}\}$ of Eqs.(3.7) we
obtain a map $T_h$ of the space
of holomorphic bundles into itself. Moreover, solutions $h=\{h_{\a\b}\}$
of Eqs.(3.7) that are close to the identity correspond to transformations
preserving topological triviality of bundles, and we obtain
the transformations
$$
\cf\ni f \ \stackrel{T_h}{\mapsto}\ f^h\in\cf .$$
In principle, by solving Eqs.(3.7) one can obtain all elements of the
group of local bijections of the space $\cf$.

\subsection{Actions of groups on the solution space}

We consider the trivial $G$-bundle $P_0$ with the transition functions
$f_0=\{\id_{U_\a\cap U_\b}\}$ and a flat (0,1)-connection $A^{0,1}=
\{A^{(\a )}\}$ on $P_0$. As it was discussed in Sect.2, for any flat
(0,1)-connection $A^{0,1}$ there exists  a 0-cochain $\psi =\{\psi_\a\}
\in\fx\subset C^0(\fu ,\fs )$ such that $A^{0,1}=\psi\bar\p\psi^{-1}=
\{\psi_\a\bar\p\psi_\a^{-1}\}$. If we denote by $\vp$ a section of the
fibration $\bar\d^0:\ \fx\to\cn$, $\bar\d^0\circ\vp=\id$, then
$\psi =\vp (A^{0,1})$, where $A^{0,1}\in\cn$, $\psi\in\fx$.
Notice that $\vp$ is a local section, i.e. we consider an open neighbourhood
of the point $A^{0,1}\in\cn$. The choice of $\vp$ is not unique, and
an element $\psi =\vp (A^{0,1})$ is defined up to an element from the group
$\fh =C^0(\fu ,\ch )$. Using the maps $\vp$ and $\eta$, we obtain a map
$$
\eta\circ\vp :\ (f_0,A^{0,1})\mapsto (f,0),
\eqno(3.8)
$$
where $f=\{f_{\a\b}\}=\{\psi^{-1}_\a\psi_\b\}$  are transition functions
of a $G$-bundle $P$.

\smallskip

Conversely, denote by $\zeta$ a local section of the fibration
$\eta :\fx\to\cf$, i.e. $\eta\circ\zeta =\id$ on an open neighbourhood
of the point $f=\{\psi^{-1}_\a\psi_\b\}\in\cf$. Of course, the choice
of a section $\zeta$ is not unique, and an element $\psi=\zeta (f)\in\fx$
is defined up to an element from the gauge group $\cg =H^0(Z,\fs )$.
Using the maps $\zeta$ and $\bar\d^0$, we obtain a map
$$
\bar\d^0\circ\zeta :\ (\tilde f, 0)\mapsto (f_0,\tilde A^{0,1}),
\eqno(3.9)$$
where $\tilde f$ is an element from an open neighbourhood of $f\in\cf$,
and $\tilde A^{0,1}$ is an element from an open neighbourhood of
$A^{0,1}\in\cn$.

\smallskip

In Sect.3.1 we have described maps $T_h:f\mapsto f^h$, where $f^h\in\cf$
if $h\in\fc$. Then, using the map (3.9) for $\tilde f=f^h$, we obtain
$$
\tilde A^{0,1}=\psi^h\bar\p(\psi^h)^{-1}=\{\psi^h_\a\bar\p
(\psi^h_\a)^{-1}\},
\eqno(3.10)$$
where $\psi^h\equiv\tilde\psi =\zeta (f^h)$ can be found from formula (3.4b).
By construction, $\tilde A^{0,1}$ satisfies Eqs.(1.2a). Thus, if we take
a ``seed" flat (0,1)-connection $A^{0,1}$ and carry out the sequence of
transformations
$$
A^{0,1}\stackrel{\vp}{\mapsto}\psi\stackrel{\eta}{\mapsto}f
\stackrel{T_h}{\mapsto}f^h\stackrel{\zeta}{\mapsto} \psi^h
\stackrel{\bar\d^0}{\mapsto}\tilde A^{0,1},
\eqno(3.11)$$
we obtain a new flat (0,1)-connection $\tilde A^{0,1}$ depending nonlocally
on $A^{0,1}$ and $h\in\fc$.

\smallskip

Let us introduce
$$
\phi (h):=\psi^h\psi^{-1}=\{\psi^h_\a\psi^{-1}_\a\}=\{\phi_\a(h)\}
\in C^0(\fu ,\fs ).
\eqno(3.12)$$
Then we have
$$
\tilde A^{0,1}=\phi (h) A^{0,1}\phi (h)^{-1} + \phi (h) \bar\p\phi (h)^{-1}=
\{\phi_\a (h) A^{(\a )}\phi_\a (h)^{-1} + \phi_\a (h) \bar\p\phi_\a (h)^{-1}\}.
\eqno(3.13)
$$ Formally, (3.13) looks like a gauge transformation. But actually
the transformation
$$
\Ad_{\phi (h)}:\ A^{0,1}\mapsto\Ad_{\phi (h)} A^{0,1}=
\phi (h) A^{0,1}\phi (h)^{-1} + \phi (h) \bar\p\phi (h)^{-1}
\eqno(3.14a)$$
defined by (3.12) consists of the sequence (3.11) of transformations
and is not a gauge transformation since $\phi_\a(h)\ne \phi_\b(h)$ on
$U_\a\cap U_\b\ne\vn$. Recall that for gauge transformations
$\Ad_g: A^{0,1}\mapsto\Ad_g A^{0,1}= gA^{0,1}g^{-1} + g\bar\p g^{-1}$
one has $g_\a =g_\b$ on $U_\a\cap U_\b$ for $g=\{g_\a\}\in\cg$.
In other words, $g=\{g_\a\}$ is a {\it globally} 
defined $G$-valued function on $Z$ and $\phi (h) =\{\phi_\a (h) \}$ 
is a collection of {\it locally} defined
$G$-valued functions $\phi_\a (h): U_\a\to G$ which are constructed by
the algorithm described above. So, to each $h\in\fc$ there corresponds
a bijective transformation $\Ad_{\phi(h)}$ of the set $\cn$, and the map
$\g : h\mapsto \Ad_{\phi (h)}$ is a homomorphism of the group $\fc$
into the group $Bij(\cn )$ of all bijections of the set $\cn$.

\smallskip

{}From formulae (2.33) and (3.12) it follows that the transformations
$\Ad_{\phi (h)}$ act on any solution of Eqs.(1.2b) by the formula
$$
\Ad_{\phi (h)}: B\mapsto\Ad_{\phi (h)} B=
\phi (h) B\phi (h)^{-1}=\{ \phi_\a (h) B^{(\a )}\phi_\a (h)^{-1}\},
\eqno(3.14b)$$
where $\phi (h)=\{ \phi_\a (h)\}$ is defined in (3.12). As is shown above,
the transformation $\Ad_{\phi (h)}$ is not a gauge transformation and
therefore $\Ad_{\phi (h)}B$ is a new solution of Eqs.(1.2b). So, we have
described a homomorphism of the group $\fc$ into the group $Bij(\ct )$
of bijective transformations of the space $\ct$ of solutions to equations
of motion
of holomorphic BF theory. The transformations (3.14) will be called the
{\it dressing transformations}. In this terminology we follow the
papers~\cite{ZS, STS, BB}, where analogous transformations were used for
constructing solutions of integrable equations.

\section{Dressing symmetries and special hBF theories}

Let now $Z$ be a Calabi-Yau $n$-manifold. This means that besides
a complex structure, on $Z$ there exist a K\"{a}hler 2-form $\o$,
a Ricci-flat K\"{a}hler metric $\mathbf g$ and a nowhere vanishing holomorphic
$(n,0)$-form $\th$. We consider a (trivial) principal $G$-bundle
$P_0$ over $Z$ and the (0,1)-component $A^{0,1}$ of a connection
1-form $A$ on $P_0$. The existence on $Z$ of a nowhere degenerate
holomorphic  $(n,0)$-form $\th$ permits one to introduce one more
class of models describing holomorphic structures on bundles over
Calabi-Yau manifolds. These models are called {\it holomorphic
$\th$BF theories} (h$\th$BF) and their action functional~\cite{Po1}
have the form
$$ S_{\mathrm{h}\th\mathrm{BF}}=\int_Z\th\wedge
\Tr(B^{0,n-2}\wedge F^{0,2}_A), \eqno(4.1) $$
where $B^{0,n-2}$ is a $\fg$-valued $(0,n-2)$-form on $Z$, and
$F^{0,2}_A$ is the $(0,2)$-component of the curvature tensor of a
connection $A$ on $P_0$. The action (4.1) leads to the following
equations of motion:
$$ \bar\p A^{0,1}+A^{0,1}\wedge A^{0,1}=0, \eqno(4.2a) $$
$$ \bar\p B^{0,n-2}+ A^{0,1}\wedge B^{0,n-2} -
(-1)^nB^{0,n-2}\wedge A^{0,1}=0.  \eqno(4.2b) $$
So, the action (4.1) provides us with a field-theoretic description
of holomorphic  structures on bundles over Calabi-Yau manifolds.

\smallskip

A description of solutions and symmetries of Eqs.(4.2) literally
reproduce a description of solutions and symmetries of field
equations (1.2) of holomorphic BF theories. In particular, the description
of flat (0,1)-connections is not changed since Eqs.(4.2a)
coincide with Eqs.(1.2a), and for a description of solutions to Eqs.(4.2b)
it is sufficient to replace $B\in\Omega^{n,n-2}(Z,\fg )$ by
$B\in\Omega^{0,n-2}(Z,\fg )$ in all formulae of Sect.2.4 and Sect.3.2.

\smallskip

One can also consider special holomorphic BF theories on twistor
spaces of self-dual 4-manifolds~\cite{Po1}. To describe these models,
we consider a Riemannian real 4-manifold $M$ with self-dual Weyl
tensor (a {\it self-dual manifold}) and the bundle $\tau : Z\to M$ of
complex structures on $M$ (the {\it twistor space} of $M$) with $\C
P^1$ as a typical fibre~\cite{Pe,AHS}. The twistor space $Z$ of a
self-dual 4-manifold $M$ is a complex 3-manifold~\cite{AHS} which is
the total space of a fibre bundle over $M$ associated with the
bundle of orthonormal frames on $M$.

\smallskip

Using the complex structures on $\C P^1\hra Z$ and $Z$, one can
split the complexified tangent bundle of $Z$ into a direct sum
$$ T^\C(Z)=T^{1,0}\oplus T^{0,1}=(Ver^{1,0}\oplus Hor^{1,0})\oplus
(Ver^{0,1}\oplus Hor^{0,1})
\eqno(4.3) $$
of subbundles of type $(1,0)$ and $(0,1)$.
Here $Ver^{0,1}$ is the distribution of vertical (0,1)-vector fields.
Analogously the complexified cotangent bundle of $Z$ is splitted into
a direct sum of subbundles $T_{1,0}$ and $T_{0,1}$.

\smallskip

Let $E^{3,3}$ be a $\fg$-valued (3,3)-form on $Z$, and $V^{0,1}$ be
an arbitrary (0,1)-vector field from the distribution $Ver^{0,1}$.
Consider a trivial $G$-bundle $P_0$ over $Z$, the (0,1)-component $A^{0,1}$
 of a connection $A$ and the (0,2)-component $F^{0,2}_A$ of the curvature
of a connection $A$ on $P_0$. Denote by  $V^{0,1}\contra A^{0,1}$
the contraction of  $V^{0,1}$ with $A^{0,1}$ and consider the
action~\cite{Po1}
$$
S_{\mathrm{hBFE}}= \int_Z\Tr[B^{3,1}\wedge
F^{0,2}_A-\g (V^{0,1}\contra A^{0,1})E^{3,3}], \eqno(4.4) $$
where $B^{3,1}\in\Omega^{3,1}(Z, \fg )$ and $\g =const$.
This action leads to the following field equations:
$$ \bar\p A^{0,1}+A^{0,1}\wedge A^{0,1}=0,\quad
\g  V^{0,1}\contra A^{0,1}=0, \eqno(4.5a) $$
$$\bar\p B^{3,1}+A^{0,1}\wedge B^{3,1} - B^{3,1}\wedge A^{0,1}= \g
V^{0,1}\contra E^{3,3}.  \eqno(4.5b) $$
Equations (4.5a) on the twistor space $Z$ of a self-dual 4-manifold
$M$ are equivalent to the self-dual Yang-Mills (SDYM) equations on
$M$~\cite{Wa, AHS}. So, the action (4.4) can be considered as an
action for SDYM models. Dressing symmetries of Eqs.(4.5a) have been
described in~\cite{Iv, Po2}. These symmetries can be reduced to
symmetries of integrable models in less than four dimensions.

\end{document}